\documentclass[twocolumn,showpacs,preprintnumbers,amsmath,amssymb]{revtex4}

\usepackage{graphicx}
\usepackage{dcolumn}
\usepackage{bm}

\begin{document}


\title{A Femtosecond Nanometer Free Electron Source}

\author{Peter Hommelhoff}
 \email{hommelhoff@stanford.edu}
\author{Yvan Sortais}%
\author{Anoush Aghajani-Talesh}%
\author{Mark A. Kasevich}%
\affiliation{%
Physics Department, Stanford University\\
Stanford, CA 94305}%

\date{July 25, 2005} 

\begin{abstract}
We report a source of free electron pulses based on a field
emission tip irradiated by a low-power femtosecond laser. The
electron pulses are shorter than 70\,fs and originate from a tip
with an emission area diameter down to 2\,nm.  Depending on the
operating regime we observe either photofield emission or optical
field emission with up to 200 electrons per pulse at a repetition
rate of 1\,GHz. This pulsed electron emitter, triggered by a
femtosecond oscillator, could serve as an efficient source for
time-resolved electron interferometry, for time-resolved
nanometric imaging and for synchrotrons.
\end{abstract}



\maketitle

Continuous electron sources based on field emission can have
emission areas down to the size of a single atom. Such spatially
resolved sources have stunning applications in surface microscopy,
to the extent that atomic scale images of surfaces are commonplace
\cite{Wiesendanger1992, Tsong1990}. Due to their brightness, field
emission electron sources are also enabling for electron
interferometry. Recently, for example, such small tips have been
used to demonstrate anti-bunching of free electrons in a
Hanbury-Brown and Twiss experiment \cite{Kiesel2002}.

On the other hand, the recent development of ultrafast pulsed
electron sources has enabled time-resolved characterization of
processes on atomic time scales. For example, the melting of a
metal has been observed with 600-fs electron pulses
\cite{Siwick2003}. Sub-femtosecond electron pulses have been used
to study the ionization dynamics of H$_2$ \cite{Niikura2002}. Fast
electron pulses are typically generated by focusing an amplified
high-power femtosecond laser beam onto a photocathode
\cite{Ihee2001} or a vapor target. In this case, the electron
emission area is given by the laser spot diameter, which is on the
order of or larger than $1\,\mu$m, much larger than the emission
area for continuous sources.

Emerging applications, such as ultrafast electron microscopy
\cite{King2005}, will require complete control over the
spatio-temporal characteristics of the emitted electrons. In this
work we realize this control through use of a low-power
femtosecond laser oscillator to trigger free electron pulses from
sharp field emission sources. Sharp tips and femtosecond lasers
have previously been combined in the context of time-resolved
scanning tunneling microscopy
\cite{Takeuchi2004,Merschdorf2002,Grafstroem02}.

\begin{figure}
\centering{
\includegraphics[width=6cm]{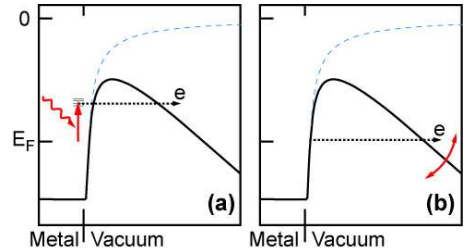}
} \caption{(color online) \label{fig1} Photofield emission (a) and
optical field emission (b) energy diagrams. In photofield emission
an electron is excited by a laser photon to an intermediate state
and then tunnels through the barrier, which is generated by a DC
voltage applied to the tip. In optical field emission, the laser
field instantaneously wiggles the barrier. If the barrier is
sufficiently thin electrons tunnel from the Fermi level. This
process is dominant for high fields. Dashed line: no field applied
on the tip.}
\end{figure}

For weak optical fields, photoemission is dominated by the
photofield effect \cite{Lee1973}, in which an initially bound
electron is promoted in energy by $\hbar \omega$ through
absorption of a single photon of frequency $\omega$ and
subsequently tunnels to the continuum [Fig.\,\ref{fig1}(a)].  Due
to the physical characteristics of the tunneling process, electron
emission is prompt with respect to the incident electric field.
For stronger optical fields, the local electric field associated
with the optical field directly modifies the tunneling potential
[optical field emission, Fig.\,\ref{fig1}(b)], again leading to
prompt electron emission. We are able to continuously tune between
the photofield and optical field emission regimes by varying the
intensity of the driving laser. These prompt mechanisms compete
with thermally induced emission, which takes place on time scales
of tens of femtoseconds to picoseconds
\cite{Fann1992,Riffe1993,Merschdorf2003}. We are able to find
operating conditions where the thermal mechanisms are negligible.

\begin{figure}
\centering{
\includegraphics[width=4.5cm]{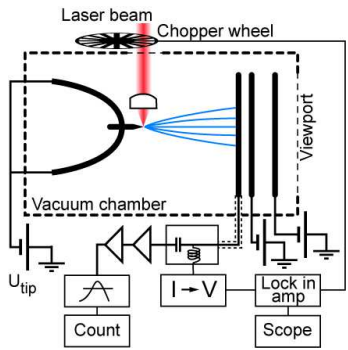}
} \caption{(color online) \label{fig2}Experimental configuration.
Sub-70 fs laser pulses hit the tip under grazing incidence at a
repetition rate of 1~GHz. Single emitted electrons are amplified
by a micro-channel plate, imaged on a phosphor screen and detected
electronically.}
\end{figure}

In our experiment, the output from a Kerr-lens mode-locked
Ti:sapphire laser is focused on a field emission tip
(Fig.\,\ref{fig2}). The laser operates at a 1 GHz repetition rate,
and produces a train of 48\,fs pulses (measured with an
interferometric autocorrelator) at a center wavelength of $\lambda
\sim$ 810\,nm with maximum average power of 600\,mW. The field
emission tip is made of electro-chemically etched $0.125\,$mm
diameter tungsten single crystal wire in the (111) orientation.
The tip is mounted in an ultra-high vacuum chamber and faces a
micro channel plate detector (MCP) located 4\,cm away from the
tip. Field emitted electrons are accelerated onto the MCP
detector. The amplified output is proximity focused on a phosphor
screen. A CCD camera records the resulting image, which reflects
the spatial distribution of photo-electrons. The time-of-arrival
of amplified photo-electrons is obtained by monitoring the MCP
bias current. At high MCP gains, we obtain spatial and temporal
single electron detection resolution.

The local electric field strength at the tip is determined by the
laser beam parameters (spot size, power, pulse duration and
polarization), and local field enhancements due to plasmon
resonances and lightning rod effects (see, for example,
\cite{Martin2001}). We focus the laser output to a $3\, \mu$m spot
size ($1/e^2$ radius) at the tip with an aspheric lens mounted
within the vacuum chamber. The propagation vector of the laser
beam is perpendicular to the tip shank, and an achromatic
half-waveplate outside the chamber is used to control the beam's
polarization. We estimate that the focusing lens ($f=7.5\,$mm)
stretches the pulses to approximately 65\,fs in the focus (see
\cite{Kempe1992,Horvath1993}), so that the peak intensity at the
tip is $3\cdot 10^{10}\,\mbox{W}/\mbox{cm}^2.$ For tungsten, the
plasmon enhancements are relatively weak, while the lightening rod
enhancement is $\sim$\,5 for a tip with a radius of curvature $r <
\lambda / 5$ \cite{Martin2001}. Thus, we estimate the maximum
electric field at the tip to be in excess of
$1\,\mbox{GV}/\mbox{m}.$

\begin{figure*}
\centering{
\includegraphics[width=12cm]{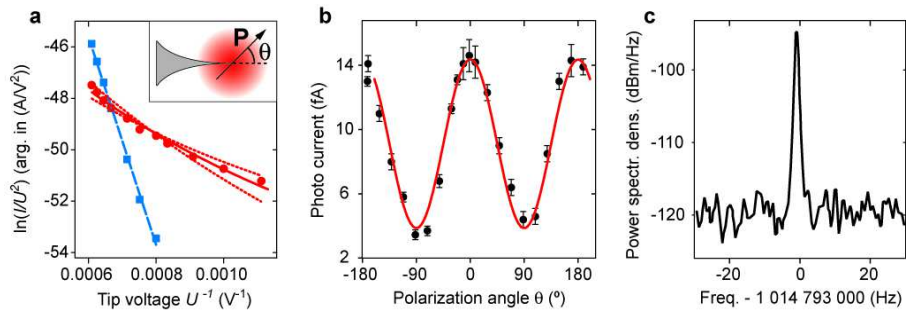}
} \caption{\label{fig3}(color online) Emission for low laser
power. (a) Fowler-Nordheim plots of the DC and photo current
(squares and circles). A fit to the DC data (dashed line) yields a
tip radius $r$ of $(134\pm 3)\,$nm. The solid line is a fit to
photofield emission current with $r= 134\,$nm and an effective
work function $\Phi = \Phi_{W} - h\nu = 3$\,eV (with laser power
$P = 260\,$mW, $\Phi_{W} = 4.5\,$eV work function of tungsten).
The presence of the laser field $F_{laser}$ at the tip further
reduces the barrier in addition to the field due to the DC voltage
applied to the tip. Therefore, we leave $F_{laser}$ as a free
parameter. The solid red line is drawn with the best fit value of
$F_{laser}=1.1\,\mbox{GV}/\mbox{m}$, and the dotted lines
represent a deviation of $\pm 25\,\%.$ With the tip's radius of
curvature much smaller than the laser wavelength we expect the
laser field to be enhanced at the tip apex by a factor of $\sim 5$
\cite{Martin2001}. By comparing the fitted $F_{laser}$ to the
maximum field calculated in the focal spot in the absence of any
material, we infer an enhancement factor of $4.1,$ in good
agreement with the expected factor. (b) Polarization dependence of
the photo-current with $U_{tip} = -1500\,$V and $P = 260\,$mW for
$r=134\,$nm. The data are well fit with a $\cos^2\theta$ on a flat
background (red line). The inset in (a) shows the definition of
the angle $\theta$ (in plane perpendicular to laser propagation
direction). (c) 1-GHz repetition rate signal measured in the
electron current. The measured line width is about 1\,Hz at
$-3\,$dBc, which corresponds to the resolution limit of the
spectrum analyzer. Error bars indicate statistical uncertainties.}
\end{figure*}

To experimentally determine the relevant emission mechanism, and
in particular to demonstrate that electron emission is prompt with
respect to the incident field, we studied emission characteristics
as a function of the DC bias voltage, laser intensity and laser
polarization. Fig.\,\ref{fig3} shows emission data taken with a
$r$ = 130 nm tip in the photofield regime.  In (a) we measure the
emitted current $I$ as a function of\ the tip bias voltage $U$,
both with and without the laser illumination. In both cases, data
are fit to the Fowler-Nordheim equation \cite{Binh1996}, which
relates the tunnel current density $j$ to the local electric field
strength $F$ and the effective work function $\Phi$:
\begin{equation} j = \frac{e^3 F^2}{8 \pi h \Phi t^2(w)}\exp
\left[ -\frac{8 \pi \sqrt{2m} \Phi^{3/2}}{3 h e F}v(w) \right].
\label{eq:FN}
\end{equation} Here, $e$ is the electron charge, $h$ Planck's
constant, $m$ the electron mass, $0.4 <v(w)<0.8$ is a slowly
varying function taking into account the image force of the
tunneling electron, $t^2(w)\approx 1$ for field emission and $w =
e^{3/2}\sqrt{F/(4 \pi \epsilon_0)}/\Phi.$ $F$ is determined
through $F = U/(kr)$ with $k=5.7$ \cite{Gomer1961PlusComment}. The
electron current $I$ is related to the current density through
$I=2\pi R^2 j,$ with $R$ the radius of the emitter area.

We use the measurements without illumination and the known work
function of tungsten to infer the tip radius $r.$ Since $r$ does
not change under laser illumination, we can then use this value to
determine the effective work function when the tip is illuminated
with the laser.  We deduce that the effective work function is
reduced by $1.5\,$eV under illumination, which corresponds to the
energy of the absorbed 810 nm laser photon.  We verified that the
value of the inferred effective work function was insensitive to
laser power for low laser power. Fig.\,\ref{fig3}(b) shows the
polarization dependence of the photocurrent, which exhibits a
$\cos^2\theta$ behavior, where $\theta$ is the angle between the
tip shank and polarization vector for the field. This is
indicative of optical excitation of surface electrons, since
translation symmetry prohibits excitation by the field component
parallel to the surface \cite{Venus83SurfSci}. Note that for
thermally increased field emission we also expect a sinusoidal
variation of the photocurrent with polarization angle. However,
from Fresnel's equations we expect that for the given tip geometry
and spot diameter the tip is heated less if the light polarization
is parallel to the tip and therefore, that the current reaches a
maximum at $\theta = 90^{\circ}$ \cite{Hadley1985}. We indeed
observed such a dependence in cw laser operation with a much
smaller modulation depth. Fig.\,\ref{fig3}(c) displays the
spectrum analysis of the electron current around the laser
repetition rate; a 30\,dB signal-to-noise ratio peak is evident at
the laser repetition rate. Taken together, these results show
that, for these parameters, the processes involved in the electron
emission are dominated by photofield emission. Thus we infer that
electron emission is prompt with respect to the laser pulse, and
rule out possible thermal emission mechanisms associated with
laser induced heating of the tip
\cite{Hadley1985,Fann1992,Riffe1993,Merschdorf2003}. For
$U_{tip}>-1300\,$V more than $98\%$ of the emitted electrons are
photo-emitted.

\begin{figure}
\centering{
\includegraphics[width=5.5cm]{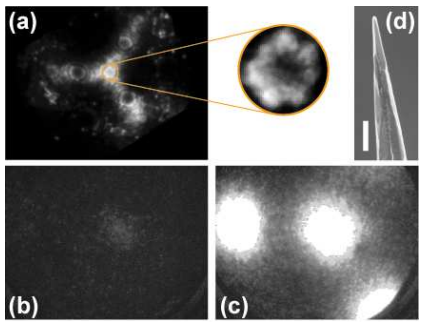}
} \caption{\label{fig4}(color online) Few atom electron source.
(a) Field ion microscope image of a 30\,nm radius tip. The zoom-in
reveals the central (111) emission area encircled by a ring of 7
atoms, spanning an area of $\lesssim$\,2\,nm diameter. (b)
DC-field emission image of a 30\,nm radius tip; only emission from
the central (111) plane is visible. (c) Same tip at the same
DC-voltage and same gain settings with laser ($P=260\,$mW). Here,
emission from the central (111) plane and neighboring \{100\}
planes saturate the MCP. (d) Scanning electron microscope image
(SEM) of a typical tip, scale bar 1\,$\mu$m.}
\end{figure}

Decreasing the tip radius leads to smaller emission planes, and
emission from a single atom is possible
\cite{Fink88,Binh1996,Fu01}. Fig.\,\ref{fig4}(a) shows a field ion
microscope (FIM) image of a 30\,nm tip. The central emission plane
is evident, consisting of a ring of 7 atoms and having an
effective area of $\sim$\,2\,nm diameter \cite{Tsong1990}.
Fig.\,\ref{fig4}(b) shows the corresponding field emission (FEM)
image at low bias voltage, without driving laser pulses. Each
grain on the image indicates the detection of an individual
electron. Emission from the central atom cluster is the dominant
contributor to the photocurrent. Finally, Fig.\,\ref{fig4}(c)
shows the same tip, at the same bias voltage, illuminated with
laser pulses. In this image, the MCP is saturated due to the more
than 100 times higher count rate. Although the count rates are
substantially higher, the basic structure of the image is
unchanged, indicating emission is coming from the sites identified
in the FIM and non-illuminated FEM images. In this regime less
than one electron is emitted per laser pulse. Due to the smallness
of the emission area, such an electron beam is well suited to
pulsed interferometer applications \cite{Kiesel2002}.

\begin{figure} \centering{
\includegraphics[width=5.5cm]{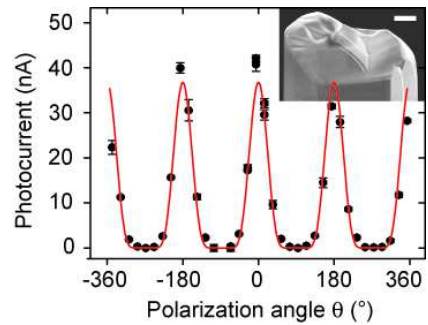}
} \caption{(color online) \label{fig5}Polarization dependence for
high laser power. Photo-current as a function of $\theta$ for a
stub as shown in the inset (SEM image, scale bar represents
500\,nm). The tip ends as a fairly flat surface facing the MCP.
Due to the sharpness of parts of the edges (radius of curvature
$\lesssim$\,100\,nm) field enhancement takes place as for sharp
tips ($P=530\,$mW, $U_{tip} = -100\,$V). The line is a fit of the
data to optical field emission (see text). Error bars indicate
statistical uncertainties.}
\end{figure}

For applications which benefit from higher currents -- but not
necessarily atomic-scale localization of the emission sites -- we
investigated emission from blunt tips. Fig.\,\ref{fig5} displays
the polarization dependence of the emission current for a tip that
ends in a flat $1\,\mu$m radius area, but which still exhibits
sharp features for field enhancement \cite{Jersch1998FieldEnh}.
Under 530\,mW illumination, time-averaged photocurrent rises to
$40\,\mbox{nA}$ and is, as before, maximal when the field is
parallel to the tip. In this case, however, the optical current
exhibits a strong nonlinearity in the field component parallel to
the tip and can be consistently fit to optical field emission
behavior. In this case \begin{equation} j = G(F_{dc} + F_{laser}
\cos\theta)^2
\exp\left[-\frac{H}{F_{dc}+F_{laser}\cos\theta}\right].
\label{eq:OFETunneling}
\end{equation} Here, $F_{laser}$ is the instantaneous absolute value of the
 laser field at the tip, $F_{dc}$ the applied dc field, and $G$ and $H$ are
constants. This expression is obtained from Eq.\,\ref{eq:FN} by
taking the local electric field to be the sum of the DC bias field
and the incident laser field. Note that this expression is
inherently time-dependent due to the presence of $F_{laser}.$
Since we measure the photocurrent time averaged over more than one
optical cycle, we fit the data in Fig.\,\ref{fig4} to
Eq.\,\ref{eq:OFETunneling} with $\cos \theta$ replaced with
$\left| \cos\theta\right|,$ assuming that electron emission is
dominated by emission at the highest field strengths. Note that
due to the exponential dependence of tunneling rate on the applied
electric field, even modest intensity changes can dramatically
alter emission characteristics. At this high laser power we find
that sharper tips ($r < 100\,$nm) experience significant
redistribution of emitting atoms, by comparing FIM images from
before and after tip illumination. On average about 200 electrons
per pulse are drawn from the tip for $\theta = 0,$ which
corresponds to an instantaneous current of $500\,\mu$A or $3.1
\cdot 10^{15}$ electrons per second. Assuming the electrons are
emitted uniformly over 65\,fs and over
the entire surface, we can set a lower limit of 
$15\,\mbox{kA}/\mbox{cm}^2$ on the instantaneous current density
and a lower limit on the invariant brightness of
$10^{13}\,\text{A}/(\text{m}^2\text{sr})$. Both values are
comparable to state-of-the-art electron pulses drawn from
photocathodes in synchrotron electron sources \cite{King2005}.

In the future, we envision the techniques demonstrated in this
work may lead to generation of sub-1 femtosecond pulses from
single-atom tips by exploiting the non-linearity in the laser-tip
interaction \cite{Baltuska2003,Niikura2002}. Likewise, this
non-linearity might enable a direct measurement of the
carrier-envelope phase of the laser pulse \cite{Paulus2003,
Apolonski04, Fortier04,Sansone2004}. Since single atom tips have
been shown to emit electrons via localized states with a lifetime
in the range of a few femtoseconds \cite{Binh1992,Yu1996}, tip
excitation with laser pulses of a similar or shorter duration in
the optical field emission regime may lead to the development of
deterministic single electron sources \cite{Zrenner2002}, which
may have important applications in quantum information science.
Finally, optimized nano-fabricated tip geometries may lead to
sources of unprecedented emission brightness.

We are indebted to Mingchang Liu and Kai Bongs for assistance in
the very early stages of this experiment and to Ralph DeVoe and
Steve Harris for discussions. This work was supported by the ARO
MURI program. P.H. thanks the Humboldt Foundation for a Lynen
Fellowship and A. A.-T. the DAAD.

\bibliographystyle{phprl}

\begin{thebibliography}{10}
\expandafter\ifx\csname url\endcsname\relax
  \def\url#1{{\tt #1}}\fi
\expandafter\ifx\csname
urlprefix\endcsname\relax\def\urlprefix{URL }\fi

\bibitem{Tsong1990}
T.~T. Tsong.
\newblock {\it Atom-probe field ion microscopy\/} (Cambridge University Press,
  Cambridge, 1990).

\bibitem{Wiesendanger1992}
R.~Wiesendanger and H.-J. G{\"u}ntherodt, eds.
\newblock {\it Scanning Tunneling Microscopy II\/} (Springer, Berlin Heidelberg
  New York, 1992).

\bibitem{Kiesel2002}
H.~Kiesel, A.~Renz, and F.~Hasselbach.
\newblock Nature {\bf 418}, 392 (2002).

\bibitem{Siwick2003}
B.~J. Siwick, J.~R. Dwyer, R.~E. Jordan, and R.~J.~D. Miller.
\newblock Science {\bf 302}, 1382 (2003).

\bibitem{Niikura2002}
H.~Niikura, F.~L{\'e}g{\'a}re, R.~Hasbani, A.~D.~Bandrauk,
M.~Yu.~Ivanov, D.~M.~Villeneuve, and P.~B.~Corkum.
\newblock Nature {\bf 417}, 917 (2002).

\bibitem{Ihee2001}
H.~Ihee, V.~A.~Lobastov, U.~M.~Gomez, B.~M.~Goodson,
R.~Srinivasan, C.-Y.~Ruan, and A.~H.~Zewail.
\newblock Science {\bf 291}, 458 (2001).

\bibitem{King2005}
W.~E.~King, G.~H.~Campbell, A.~Frank, B.~Reed, J.~F.~Schmerge,
B.~J.~Siwick, B.~C.~Stuart, and P.~M.~Weber.
\newblock J.~Appl.~Phys. {\bf 97}, 111101 (2005).

\bibitem{Takeuchi2004}
O.~Takeuchi, M.~Aoyama, R.~Oshima, Y.~Okada, H.~Oigawa, N.~Sano,
H.~Shigekawa, R.~Mota, and M.~Yamashita.
\newblock Appl.~Phys.~Lett. {\bf 85}, 3268 (2004).

\bibitem{Merschdorf2002}
M.~Merschdorf, W.~Pfeiffer, A.~Thon, and G.~Gerber.
\newblock Appl.~Phys.~Lett. {\bf 81}, 286 (2002).

\bibitem{Grafstroem02}
S.~Grafstr{\"o}m.
\newblock J.~Appl.~Phys. {\bf 91}, 1717 (2002).

\bibitem{Lee1973}
M.~J.~G. Lee.
\newblock Phys.~Rev.~Lett. {\bf 30}, 1193 (1973).

\bibitem{Fann1992}
W.~S. Fann, R.~Storz, H.~W.~K. Tom, and J.~Bokor.
\newblock Phys.~Rev.~B {\bf 46}, 13592 (1992).

\bibitem{Riffe1993}
D.~M.~Riffe, X.~Y.~Wang, M.~C.~Downer, D.~L.~Fisher, T.~Tajima,
and J.~L.~Erskine.
\newblock J.~Opt.~Soc.~Am.~B {\bf 10}, 1424 (1993).

\bibitem{Merschdorf2003}
M.~Merschdorf, W.~Pfeiffer, S.~Voll, and G.~Gerber.
\newblock Phys.~Rev.~B {\bf 68}, 155416 (2003).

\bibitem{Martin2001}
Y.~C. Martin, H.~F. Hamann, and H.~K. Wickramasinghe.
\newblock J.~Appl.~Phys. {\bf 89}, 5774 (2001).

\bibitem{Kempe1992}
M.~Kempe, U.~Stamm, B.~Wilhelmi, and W.~Rudolph.
\newblock J.~Opt.~Soc.~Am.~B {\bf 9}, 1158 (1992).

\bibitem{Horvath1993}
Z.~I. Horv{\'a}th and Z.~Bor.
\newblock Optics~Comm. {\bf 100}, 6 (1993).

\bibitem{Binh1996}
See, for example, V.~T. Binh, N.~Garcia, and S.~T. Purcell.
\newblock Adv. Imag. Elect. Phys. {\bf 95}, 63 (1996).

\bibitem{Gomer1961PlusComment}
We determine $k$ through an iterative routine, as described in
R.~Gomer.
\newblock {\it Field emission and field
  ionization\/} (Harvard University Press, Cambridge, Massachusetts, 1961).

\bibitem{Venus83SurfSci}
D.~Venus and M.~J.~G. Lee.
\newblock Surface Science {\bf 125}, 452 (1983).

\bibitem{Hadley1985}
K.~W. Hadley, P.~J. Donders, and M.~J.~G. Lee.
\newblock J.~Appl.~Phys. {\bf 57}, 2617 (1985).

\bibitem{Fink88}
H.-W. Fink.
\newblock Physica Scripta {\bf 38}, 260 (1988).

\bibitem{Fu01}
T.-Y. Fu, L.-C. Cheng, C.-H. Nien, and T.~T. Tsong.
\newblock Phys.~Rev.~B {\bf 64}, 113401 (2001).

\bibitem{Jersch1998FieldEnh}
J.~Jersch, F.~Demming, L.~J. Hildenhagen, and K.~Dickmann.
\newblock Appl.~Phys.~A {\bf 66}, 29 (1998).

\bibitem{Baltuska2003}
A.~Baltu{\u s}ka, Th.~Udem, M.~Uiberacker, M.~Hentschel,
E.~Goulielmakis, Ch.~Gohle, R.~Holzwarth, V.~S.~Yakovlev,
A.~Scrinzi, T.~W.~H{\"a}nsch, and F.~Krausz.
\newblock Nature {\bf 421}, 611 (2003).

\bibitem{Paulus2003}
G.~G.~Paulus, F.~Lindner, H.~Walther, A.~Balt{\u u}ska,
E.~Goulielmakis, M.~Lezius, and F.~Krausz.
\newblock Phys.~Rev.~Lett. {\bf 91}, 253004 (2003).

\bibitem{Apolonski04}
A.~Apolonski, P.~Dombi, G.~G.~Paulus, M.~Kakehata, R.~Holzwarth,
Th.~Udem, Ch.~Lemell, K.~Torizuka, J.~Burgd{\"o}rfer,
T.~W.~H{\"a}nsch, and F. Krausz.
\newblock Phys.~Rev.~Lett. {\bf 92}, 073902 (2004).

\bibitem{Fortier04}
T.~M.~Fortier, P.~A.~Roos , D.~J.~Jones, S.~T.~Cundiff,
R.~D.~R.~Bhat, and J.~E.~Sipe.
\newblock Phys.~Rev.~Lett. {\bf 92}, 147403 (2004).

\bibitem{Sansone2004}
G.~Sansone, C.~Vozzi, S.~Stagira, M.~Pascolini, L. Poletto,
P.~Villoresi, G.~Tondello, S.~DeSilvestri, and M.~Nisoli.
\newblock Phys.~Rev.~Lett. {\bf 92}, 113904 (2004).

\bibitem{Binh1992}
V.~T. Binh, S.~T. Purcell, N.~Garcia, and J.~Doglioni.
\newblock Phys.~Rev.~Lett. {\bf 69}, 2527 (1992).

\bibitem{Yu1996}
M.~L.~Yu, N.~D.~Lang, B.~W.~Hussey, T.~H.~P.~Chang, and
W.~A.~Mackie.
\newblock Phys.~Rev.~Lett. {\bf 77}, 1636 (1996).

\bibitem{Zrenner2002}
A.~Zrenner, E.~Beham, S.~Stufler, F.~Findeis, M.~Bichler, and
G.~Abstreiter.
\newblock Nature {\bf 418}, 612 (2002).

\end{thebibliography}

\end{document}